\documentclass[10pt,apj,tighten,iop]{emulateapj}

\usepackage{graphicx}
\usepackage{amsmath}   
\usepackage{tabulary}
\usepackage{epsfig}
\usepackage{amssymb}
\usepackage{multirow}
\usepackage{appendix}
\usepackage{natbib}
\usepackage{lineno}
\usepackage{wasysym}
\usepackage{color}         
\usepackage{enumerate}
\usepackage{tikz}

\usetikzlibrary{arrows,shapes,calc}

\usepackage[pdftitle={Modeling the Orbital Sampling Effect of Extrasolar Moons}, colorlinks = true, breaklinks = true,
  citecolor = blue, linkcolor = blue, urlcolor = blue, pdfauthor = {Ren\'{e} Heller}]{hyperref}
\shorttitle{Modeling the OSE of Exomoons}

\submitted{Published in The Astrophysical Journal 820:88 (11pp), 2016 April 1}

\shortauthors{Heller, Hippke \& Jackson}

\newcommand{\kepler}{\emph{Kepler}}
\newcommand{\kerplertwotwoninecperiod}{16.968618\,d}

\begin{document}

\title{MODELING THE ORBITAL SAMPLING EFFECT OF EXTRASOLAR MOONS}

\author{Ren\'{e} Heller}
\affil{Max Planck Institute for Solar System Research, Justus-von-Liebig-Weg 3, 37077 G\"ottingen, Germany; \href{mailto:heller@mps.mpg.de}{heller@mps.mpg.de}}

\author{Michael Hippke}
\affil{Luiter Stra{\ss}e 21b, 47506 Neukirchen-Vluyn, Germany; \href{mailto:hippke@ifda.eu}{hippke@ifda.eu}}

\author{Brian Jackson}
\affil{Carnegie Department of Terrestrial Magnetism, 5241 Broad Branch Road, NW, Washington, DC 20015, USA}
\affil{Department of Physics, Boise State University, Boise, ID 83725-1570, USA; \href{mailto:bjackson@boisestate.edu}{bjackson@boisestate.edu}}

\begin{abstract}
The orbital sampling effect (OSE) appears in phase-folded transit light curves of extrasolar planets with moons. Analytical OSE models have hitherto neglected stellar limb darkening and non-zero transit impact parameters and assumed that the moon is on a circular, co-planar orbit around the planet. Here, we present an analytical OSE model for eccentric moon orbits, which we implement in a numerical simulator with stellar limb darkening that allows for arbitrary transit impact parameters. We also describe and publicly release a fully numerical OSE simulator ({\tt PyOSE}) that can model arbitrary inclinations of the transiting moon orbit. Both our analytical solution for the OSE and {\tt PyOSE} can be used to search for exomoons in long-term stellar light curves such as those by \textit{Kepler} and the upcoming \textit{PLATO} mission. Our updated OSE model offers an independent method for the verification of possible future exomoon claims via transit timing variations and transit duration variations. Photometrically quiet K and M dwarf stars are particularly promising targets for an exomoon discovery using the OSE.
\end{abstract}

\keywords{instrumentation: photometers -- methods: data analysis -- methods: analytical --  methods: observational -- methods: statistical -- planets and satellites: detection}

\section{Context and Motivation}
\label{sec:context}

The race toward the first detection of an extrasolar moon picks up pace. While the first attempts to detect exomoons were byproducts of planet-targeted observations \citep{2001ApJ...552..699B,2007A&A...476.1347P,2010MNRAS.407.2625M}, high-accuracy space-based observations of thousands of transiting exoplanets and planet candidates by \textit{Kepler} \citep{1997ASPC..119..153B,2013ApJS..204...24B} now allow for dedicated exomoon searches \citep{2012ApJ...750..115K,2013A&A...553A..17S,2015ApJ...806...51H}. Upcoming data from the European Space Agency's \textit{CHEOPS} and \textit{PLATO} space missions will provide further promising avenues toward an exomoon detection \citep{2015ApJ...810...29H,2015PASP..127.1084S}.

Exomoon detections will be highly valuable for our understanding of the origin and fate of planetary systems because they probe the substructures of planet formation that is not accessible through planet observations alone. As for the solar system, moons provide key insights into the formation of Earth \citep[by a giant collision;][]{1976LPI.....7..120C,2012Sci...338.1052C}, into the temperature distributions within the circumplanetary accretion disks of Jupiter and Saturn \citep{1974Icar...21..248P,2002AJ....124.3404C,2010ApJ...714.1052S,2015A&A...578A..19H}, and into the cause of Uranus' tilted spin axis \citep[by gradual collisional tilting;][]{2012Icar..219..737M}. As of today, no moon has been confirmed around a planet beyond the solar system. Hence, exoplanetary science suffers from a fundamental lack of knowledge about the fine structure of planetary systems.

\begin{figure}[t]
   \centering
   \scalebox{0.447}{\includegraphics{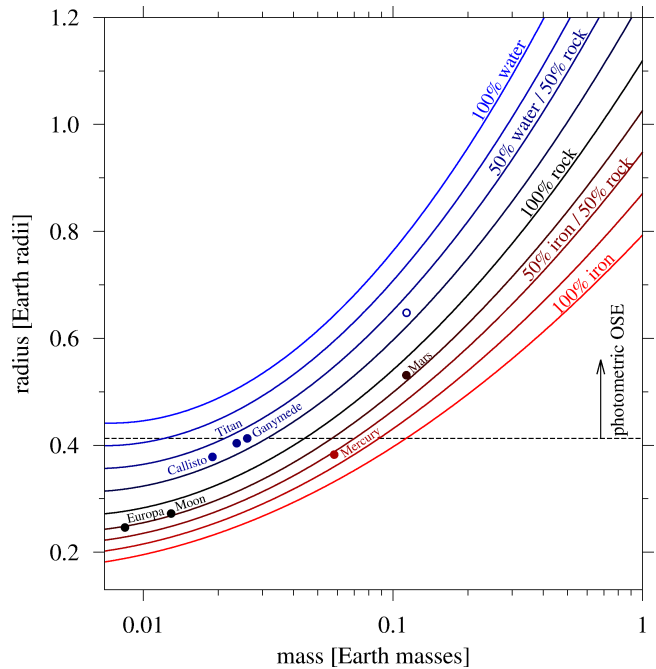}}
   \caption{Prospective detection thresholds of the photometric OSE (horizontal dashed line) for moons around a photometrically quiet star after $\gtrsim30$ transits \citep[as per][Fig.~11 therein]{2014ApJ...787...14H}. Note that the photometric OSE is sensitive to an exomoon's radius but not to its mass. Hence, large (potentially low-density) moons are the most promising targets. The open circle denotes a Mars-mass moon with a Ganymede-like composition as predicted by \citet{2015ApJ...806..181H}. The mass-radius relationship for moons of various compositions is according to \citet{2007ApJ...659.1661F}.}
   \label{fig:massradius}
 \end{figure}

\citet{2014ApJ...787...14H} recently identified a new exomoon signature in planetary transit light curves, which occurs due to the additional darkening of the star by a transiting moon. This phenomenon, which we refer to as the photometric orbital sampling effect (OSE), occurs in phase-folded transit light curves, because an exomoon's sky-projected position with respect to its host planet is variable in subsequent transits but is statistically predictable for a large number of transits ($N\gtrsim10$).\footnote{The OSE appears in three flavors \citep[as per][]{2014ApJ...787...14H}, one of which is the photometric OSE, which we focus on in this paper. The other two manifestations of the OSE appear in the planetary transit timing variations (TTV-OSE) and transit duration variations (TDV-OSE).} The photometric OSE is not sensitive to the satellite mass ($M_{\rm s}$), but it is very sensitive to its radius ($R_{\rm s}$). This makes the photometric OSE particularly sensitive to low-density, water-rich moons like the three most massive moons in the solar system, Ganymede and Callisto around Jupiter and Titan around Saturn. These moons would hardly be detectable in the available \textit{Kepler} data by combined TTV and TDV measurements, which are sensitive to roughly Earth-mass moons \citep[][]{2013A&A...553A..17S}. The most advanced exomoon search as of today, the ``Hunt for Exomoons with Kepler'' \citep[HEK;][]{2012ApJ...750..115K}, using a photodynamical model, achieves much better detection limits down to a few Ganymede masses, depending on the mass of the host planet, amongst other parameters \citep{2015ApJ...813...14K}. However, if giant planets were able to give their fully fledged, icy moon systems a piggyback ride to $\lesssim1$\,AU, where super-Jovian planets are abundant in radial velocity survey data \citep{2013ApJ...767L..24D,2015A&A...578A..19H}, then the photometric OSE could be a promising method to find these exomoons. The mass-radius diagram in Figure~\ref{fig:massradius} illustrates the detection threshold for moons transiting photometrically quiet M stars. Note that Mars-mass moons with a water-rock composition similar to Ganymede, Callisto, and Titan are significantly larger than Mars (see the open circle). Hence, they could be detectable around photometrically quiet stars. These moons are predicted to form frequently around the most massive super-Jovian planets \citep{2015ApJ...806..181H}.

 \begin{figure}[t]
   \centering
   \scalebox{0.17}{\includegraphics{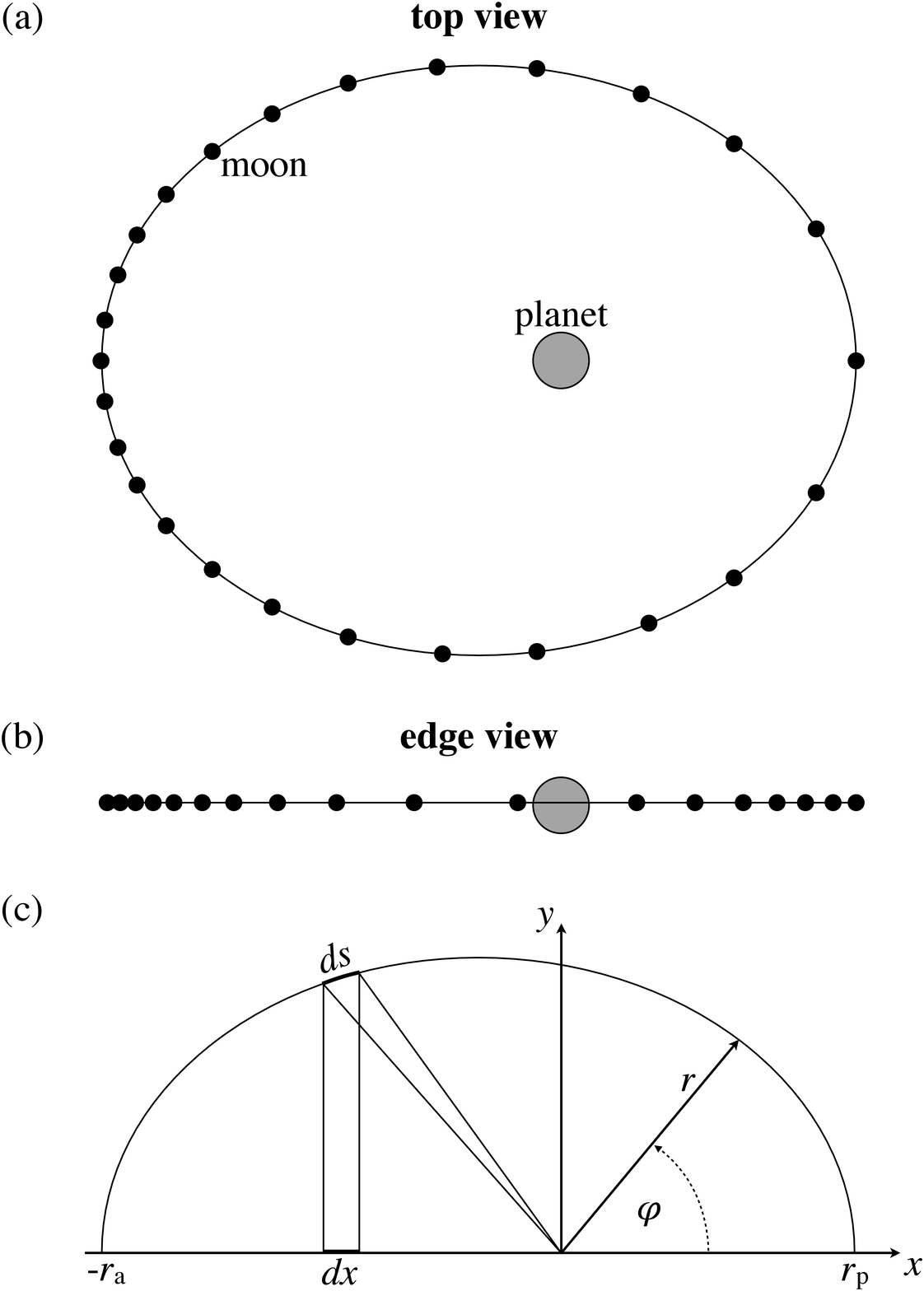}}
   \caption{Construction of an exomoon's orbital sampling frequency $\mathcal{P}_{\rm s}(x)$ for elliptical orbits. \textbf{(a)} The moon's orbital position around the planet is measured with a constant sampling frequency, or frame rate. \textbf{(b)} The moon's variable orbital velocity yields an asymmetric probability density with respect to the planet. \textbf{(c)} The probability density can be derived as $P_{\rm s}(x)=ds(x)/dx$. The special case of a circular orbit is given in Figure~1 of \citep{2014ApJ...787...14H}.}
   \label{fig:sampling}
 \end{figure}

\citet{2015ApJ...806...51H} searched for the photometric OSE in the archival data of the \textit{Kepler} telescope and found indications for an OSE-like signal in the combined transit light curves of hundreds of planets and planet candidates with orbital periods $>35$\,d. The original description of the OSE by \citet{2014ApJ...787...14H} was purely analytical and thereby fast, but it made several simplifications: stellar limb darkening was neglected; the planet--moon orbit was assumed to be non-eccentric ($e=0$) and non-inclined ($i=0^\circ$) with respect to the circumstellar orbit; the planet--moon system was assumed to transit the star along the stellar diameter; that is, the transit impact parameter ($\mathfrak{b}$) was set to 0. Consequently, this model was not broadly applicable. \citet{2015ApJ...806...51H} performed a mostly numerical study that did include non-circular planet--moon orbits and arbitrary inclinations, but he did not explore a wide parameter range of the OSE.

We here first derive a novel equation for the orbital sampling frequency $\mathcal{P}_{\rm s}(x)$ of exomoons on eccentric orbits ($e\geq0$; Section~\ref{sub:P_eccentric}). We then incorporate our formula into a numerical simulator that models the phase-folded transits of exoplanets with moons in front of stars with limb darkening (Section~\ref{sub:OSEmodel}). Our simulations are compared to real \textit{Kepler} data. We also perform purely numerical simulations for a wide range of inclined planet--moon orbits ($i\geq0$), planetary impact parameters ($\mathfrak{b}\geq0$), satellite radii ($R_{\rm s}$), and semimajor axes of the satellite's orbit around the planet ($a$; Section~\ref{sec:numerical}).

\section{A dynamical model for the photometric OSE}
\label{sec:dynamical}

\subsection{Sampling Frequency for Eccentric Exomoon Orbits}
\label{sub:P_eccentric}

In \citet{2014ApJ...787...14H}, the OSE was described for circular moon orbits only. Given that the eccentricities of the 10 largest moons in the solar system are all smaller than that of the Earth's Moon ($\approx0.055$), this approximation seems appropriate. However, the architectures and physical properties of extrasolar planetary systems turned out to be very different from the solar system planets, as we recall the discoveries of terrestrial planets around pulsars \citep{1992Natur.355..145W}, Jupiter-mass planets in extremely short-period orbits \citep{1995Natur.378..355M}, circumbinary planets \citep{2011Sci...333.1602D}, and a large population of super-Earths with short orbital periods \citep{2013ApJS..204...24B}. Hence, astronomers should not be surprised if they were to find that at least some exomoons follow significantly eccentric orbits. Eccentricities may, for example, be forced by gravitational interaction with the star \citep{2012A&A...545L...8H,2016ApJ...817...18S}, other planets \citep{2007Sci...318..244C,2013ApJ...769L..14G,2013ApJ...775L..44P}, or even the host planet \citep{1963MNRAS.126..257G}. Moreover, an analytical solution is always worthwhile as it can give a deeper understanding of the underlying physical processes that generate an observational feature.

We here derive the orbital sampling probability function for exomoons on eccentric circumplanetary orbits. Figure~\ref{fig:sampling}(a) shows the moon's circumplanetary orbit in a plane perpendicular to the reader's line of sight (solid line). If an observer were to take pictures of the moon's orbital position with a constant sampling rate, then the individual moon pictures would be distributed in a manner similar to the one depicted in Figure~\ref{fig:sampling}(a). Around periapsis, the moon's Keplerian velocity ($v$) is higher than around apoapsis. Thus, the probability of observing the moon, e.g. during a common stellar transit with its host planet, at a given orbital position is asymmetric with respect to the sky-projected distance to the planet. This fact is visualized in Figure~\ref{fig:sampling}(b), where the planet--moon system is sampled from a co-planar perspective. Due to the projection effect, the moon(s) pile(s) up toward the edges of the projected major axis, but the probability distribution to the left of the planet is different from what it looks to the right of the planet. Figure~\ref{fig:sampling}(c) shows how we construct the satellite's probability density $\mathcal{P}_{\rm s}(x)$ as a function of its projected distance ($x$) to the planet.

In its most general form, $\mathcal{P}_{\rm s}(x)$ describes what fraction of its orbital period ($P_{\rm ps}$) the satellite spends in an infinitely small interval ($dx$) on its sky-projected orbit along the $x$-axis (see Figure~\ref{fig:sampling}(c)). Hence, it is given as

\begin{equation}\label{eq:Ps_general}
\mathcal{P}_{\rm s}(x) \propto \frac{1}{P_{\rm ps}} \frac{dt}{dx} \ ,
\end{equation}

\noindent
where $dt$ is an infinitesimal change in time. With $ds$ as an infinitely small interval on the moon's elliptical orbit, the Keplerian velocity is $v(x)=ds(x)/dt$, hence

\begin{equation}\label{eq:Ps_path}
\mathcal{P}_{\rm s}(x) \propto \frac{1}{P_{\rm ps} v(x)} \frac{ds(x)}{dx} \ ,
\end{equation}

\noindent
and the challenge is then in finding $ds(x)/dx$. From Figure~\ref{fig:sampling}(a) we learn that $ds(x)=r(x)d{\varphi}(x)$. We use the parameterization of a Keplerian orbit

\begin{equation}\label{eq:Keplerian}
r(\varphi) = \frac{a(1-e^2)}{1+e\cos(\varphi)}
\end{equation}

\noindent
and solve it for ${\varphi}(x)$ via

\begin{align}\label{eq:varphi}\nonumber
& x(\varphi) \ = \ r(\varphi)\cos(\varphi) = \frac{a(1-e^2)}{\left(\frac{\displaystyle 1}{\displaystyle \cos(\varphi)}+e\right)}\\
\Leftrightarrow \ \ \ \ \ \ & \varphi(x) = \ \arccos\left( \left[ \frac{a}{x}(1-e^2) -e \right]^{-1} \right) \ .
\end{align}

\noindent
Equation~(\ref{eq:Ps_path}) then becomes

\begin{align}\label{eq:Ps_x}
\mathcal{P}_{\rm s}(x) \propto & \frac{r(x)}{P_{\rm ps} v(x)} \frac{d\varphi(x)}{dx} \\ \nonumber
= & \frac{r(x)}{P_{\rm ps} v(x)} \underbrace{ \frac{d}{dx} \arccos\left( \left[ \frac{a}{x}(1-e^2) -e \right]^{-1} \right) }
\end{align}

\vspace{-.4cm}

\begin{tikzpicture}[overlay]
\draw[thick] (1.324,-1.6) to [out=90,in=-90] (1.2,-1) to [out=90,in=-90] (5.255,-0.05);
\end{tikzpicture}

\begin{equation}\nonumber
\hspace{1cm}Ê= A \left( \sqrt{1-\frac{1}{\displaystyle \left(\frac{A}{x}-e\right)^2}} \left(\frac{A}{x}-e\right)^2 x^2 \right)^{-1} \ ,
\end{equation}

\vspace{.4cm}

\noindent
where we introduced $A{\equiv}a(1-e^2)$. In Equation~(\ref{eq:Ps_x}), $r(x)$ can be derived by inserting Equation~(\ref{eq:varphi}) into (\ref{eq:Keplerian}), hence

\begin{equation}\label{eq:r_x}
r(x) = \frac{A}{1-e\left(\frac{\displaystyle A}{\displaystyle x}-e\right)^{-1}} \ ,
\end{equation}

\noindent
and $v(x)$ in Equation~(\ref{eq:Ps_x}) is given by

\begin{equation}\label{eq:v_x}
v(r) = \sqrt{\mu \left(\frac{2}{r(x)} - \frac{1}{a} \right)} \ ,
\end{equation}

\noindent
where $\mu=G(M_{\rm p}+M_{\rm s})$ and $G$ is Newton's gravitational constant.

So far, we assumed the line of sight to be parallel to the $y$-axis in Figure~(\ref{fig:sampling}), that is, along the orbital semiminor axis ($b=a(1-e^2)^{1/2}$). Of course, the moon orbit can be rotated in the $x$-$y$ plane by an angle $\omega$. We can imagine that once $\omega=90^\circ$, the observer samples the moon orbit according to $\mathcal{P}_{\rm s}(y)$. In this case, the sky-projected orbit appears symmetric with an apparent radius ($\tilde{a}(e,\omega)$) equal to $b$. In general, we have

\begin{align}\nonumber
\tilde{a}(e,\omega) = & \ \sqrt{{\Big (}a\cos(\omega){\Big )}^2 + {\Big (}b\sin(\omega){\Big )}^2}\\
= & \ a\sqrt{\cos(\omega)^2 + (1-e)\sin(\omega)^2} \ .
\end{align}

\noindent
We then replace $e$ with $\tilde{e}=e\cos(\omega)$ and $A$ with $\tilde{A}=\tilde{a}(1-\tilde{e}^2)$ in Equations~(\ref{eq:Ps_x})--(\ref{eq:v_x}) to obtain $\mathcal{P}_{\rm s}(x)$ for arbitrary orientations $\omega$.

At last, we can swap the proportionality sign in Equation~(\ref{eq:Ps_x}) with an equality sign by normalizing

\begin{equation}\label{eq:norm}
\int_{-r_{\rm a}}^{+r_{\rm p}} dxÊ\ \mathcal{P}_{\rm s}(x) \equiv 1 \ ,
\end{equation}

\noindent
where $r_{\rm a}$ and $r_{\rm p}$ are the moon's orbital radii at apoapsis and periapsis, respectively (see Figure~\ref{fig:sampling}(c)). We evaluate this integral numerically for arbitrary $a$, $e$, and $\omega$ and find an approximate solution

\begin{align}\label{eq:Ps_x_final}
\mathcal{P}_{\rm s}(x) = & 2(1+e^4) \left( \frac{a}{\tilde{a}(e,\omega)} \right)^2 \frac{\tilde{r}(x)}{P_{\rm ps} \tilde{v}(x)} \\ \nonumber
 & \times \tilde{A} \left( \sqrt{1-\frac{1}{\displaystyle \left(\frac{\tilde{A}}{x}-\tilde{e}\right)^2}} \left(\frac{\tilde{A}}{x}-\tilde{e}\right)^2 x^2 \right)^{-1} \ ,
\end{align}

\noindent
where $\tilde{r}(x)$ and $\tilde{v}(x)$ refer to Equations~(\ref{eq:r_x}) and (\ref{eq:v_x}), respectively, but swapping $e$ for $\tilde{e}$ and $A$ for $\tilde{A}$.\footnote{For $e=0$, we have $\tilde{e}=0$, $\tilde{a}=a$, $\tilde{A}=a$, $\tilde{r}=a$, and $\tilde{v}=(\mu/a)^{1/2}$. Hence, setting $e=0$ in Equation~(\ref{eq:Ps_x_final}) and using $P_{\rm ps}=2{\pi}a(a/\mu)^{1/2}$ we obtain the circular case $\mathcal{P}_{\rm s}(x)=1/\left({\pi}a[1-(x/a)^2]^{1/2}\right)$ derived in \citet[][Equation~(4) therein]{2014ApJ...787...14H}.} The error in Equation~\ref{eq:Ps_x_final} is $\ll1\,\%$ for $e<0.9$.

\begin{figure}[t]
   \centering
   \scalebox{0.60}{\includegraphics{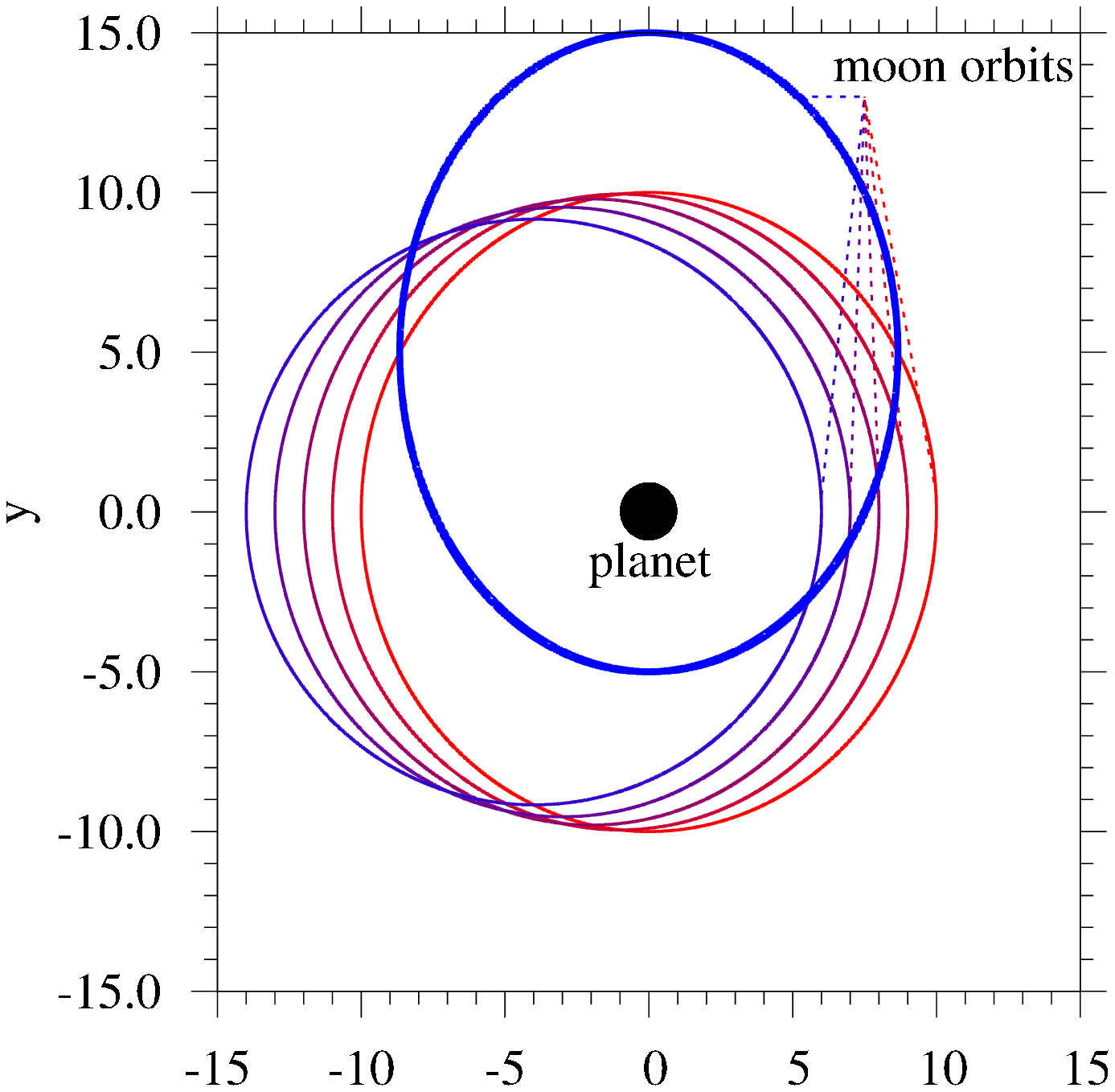}}\\
   \vspace{0.2cm}
   \scalebox{0.57}{\includegraphics{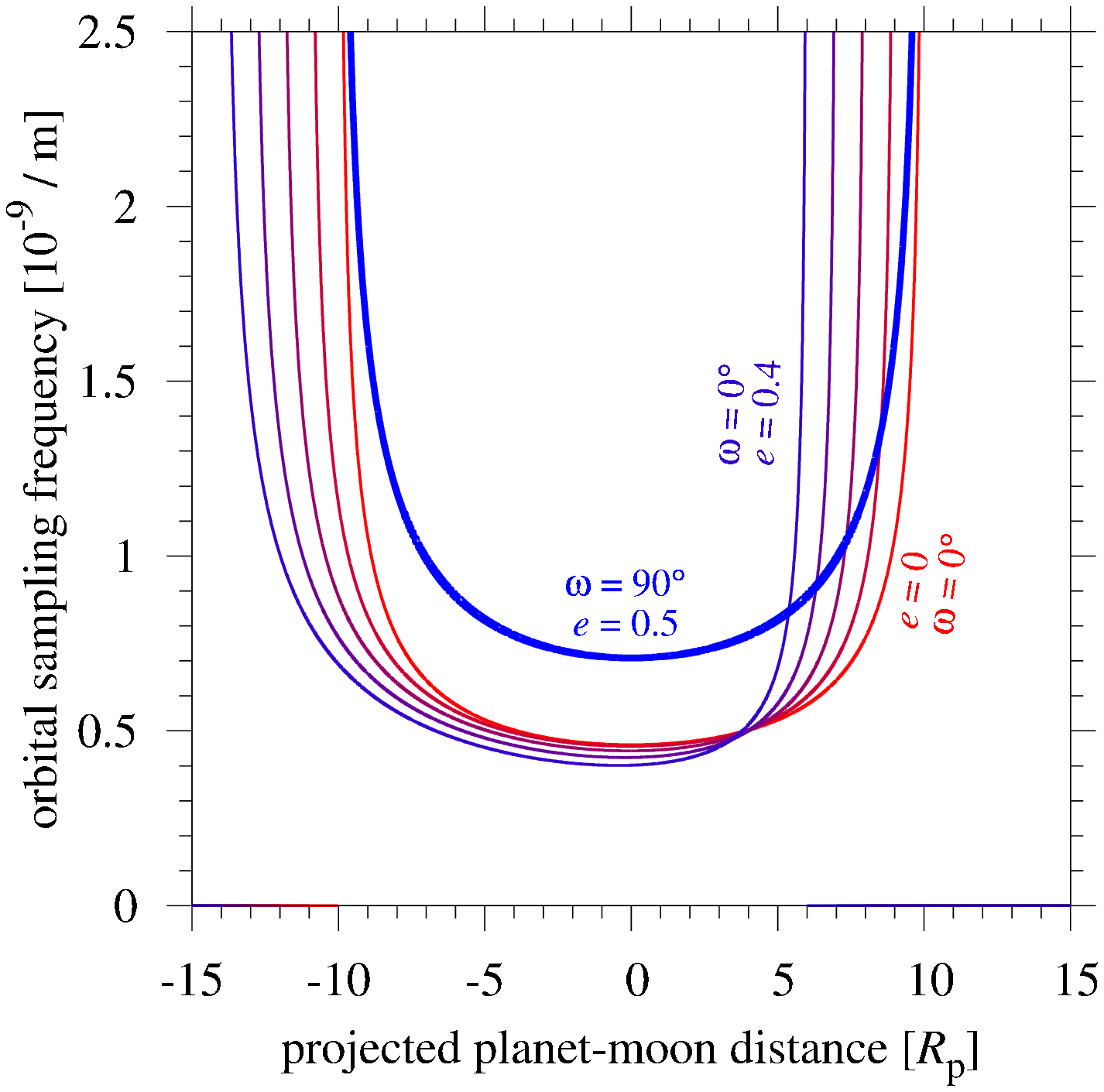}}
   \caption{Different orbital eccentricities ($e$) in the planet--moon system (upper panel) cause different orbital sampling frequencies (lower panel). Six different values are shown: $e\in\{0, 0.1, 0.2, 0.3, 0.4, 0.5\}$.}
   \label{fig:ellipse}
\end{figure}

In Figure~\ref{fig:ellipse}, we plot Equation~(\ref{eq:Ps_x_final}) for a moon orbiting a planet on various eccentric orbits, all of which have a semimajor axis of 10 planetary radii ($R_{\rm p}$). The upper panel depicts the orbital geometries and the lower panel shows $\mathcal{P}_{\rm s}(x)$ in each case. Four cases with eccentricities between 0 and 0.4 assume $\omega=0$ as presented in Figure~\ref{fig:sampling}, and the $e=0.5$ case is rotated by $\omega=90^\circ$. Note that in the latter scenario, where the semiminor axis $b$ is substantially smaller than $a$ and the line of sight is parallel to $a$, $\mathcal{P}_{\rm s}(x)$ is significantly higher at a given planetary separation because the moon occupies a smaller circumplanetary region along the $x$-axis. This means that the OSE becomes particularly prominent in the phase-folded light curve of eccentric moon systems with $\omega$ close to $90^\circ$ or $270^\circ$. On the other hand, it becomes relatively weak (or ``smeared'') for $\omega$ near $0^\circ$ or $180^\circ$.

\subsection{The Photometric OSE with Limb Darkening}
\label{sub:OSEmodel}

 \begin{figure*}[t]
   \centering
   \includegraphics[angle=-0, width=\linewidth]{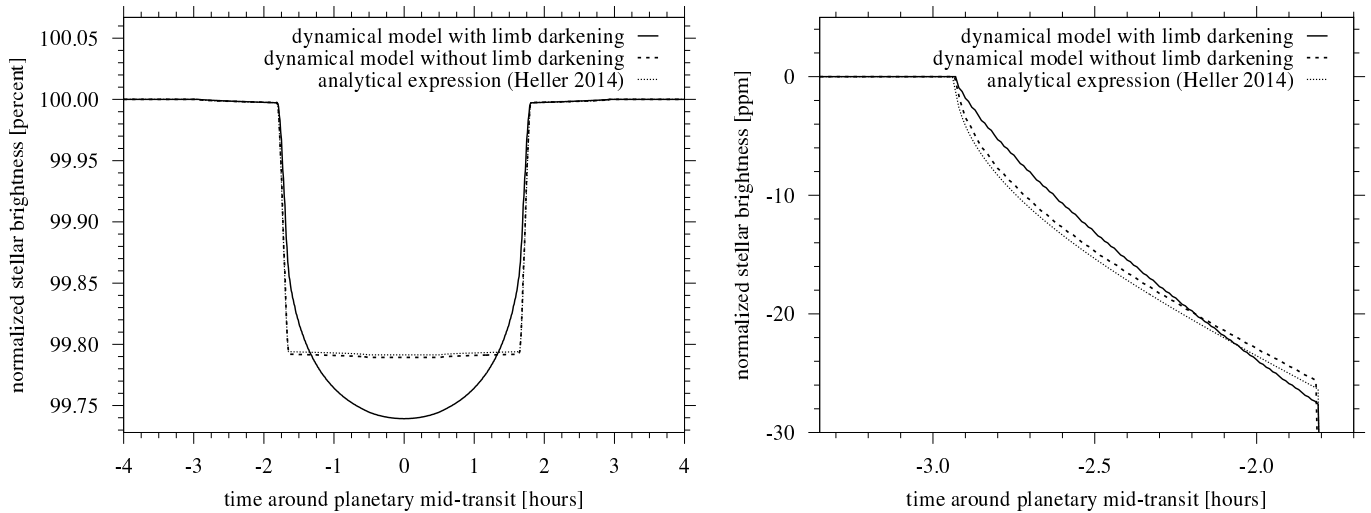}
   \caption{Comparison of the dynamical photometric OSE model with limb darkening (solid line), the dynamic photometric OSE model without limb darkening (dashed line), and the analytical expression for a non-limb-darkened star as per Eq. (6) in \citet{2014ApJ...787...14H}. \textit{Left:} The OSE is hardly visible as a the slight brightness decrease in the wings of the planetary transit. \textit{Right:} A zoom into the ingress of the moon. The dynamic model without limb darkening reproduces the analytical prediction very well. However, only the dynamic limb darkening model will be useful for fitting real observations. This hypothetical star--planet system is similar to the KOI\,255.01 system, except that we here assume $\mathfrak{b}=0$. The toy moon is as large as Ganymede ($0.41\,R_\oplus$) and the planet--moon orbit is $15.47\,R_{\rm p}$ wide, equivalent to Ganymede's orbit around Jupiter.}
   \label{fig:limb_255.01}
 \end{figure*}

\subsubsection{Dynamical Simulations}
\label{subsub:dynamical}

We built a numerical model to simulate the stellar transit of an exoplanet for arbitrary probability functions $\mathcal{P}_{\rm s}(x)$ , including multiple functions in the case of multi-moon systems. As an update to \citet{2014ApJ...787...14H}, our simulator now considers stellar limb darkening and the transit impact parameter can be varied. In comparison to \citet[][Section~2.2.2 therein]{2014ApJ...787...14H}, where an analytical solution for the actual light curve (the stellar brightness $B_{\rm OSE}^{(n)}$ due to a transiting planet with $n$ exomoons) without stellar limb darkening has been derived, we do not derive an analytical solution for the light curve of the OSE with limb darkening. Instead, we simulate the OSE by generating a computer model of a limb-darkened stellar disk, approximated as a circle touching the the edges of a square sized $1\,000\times1\,000$ pixels. Then we let the planet and the probability function(s) of the moon(s) transit. In the following, we refer to this approach as our ``dynamical OSE model''.

Once $\mathcal{P}_{\rm s}(x)$ enters the stellar disk, we multiply the stellar intensity ($I_\star^{\rm px}$) in any pixel that is covered with the probability density $\mathcal{P}_{\rm s}^{\rm px}(x)$ in this pixel, so that $dx{\times}I_\star^{\rm px}{\times}\mathcal{P}_{\rm s}^{\rm px}(x)$ is the relative amount of stellar brightness that is obscured in this pixel. In this notation, $dx$ corresponds to the pixel width. Planet--moon eclipses are automatically taken into account, because the planet is simulated as a black circle. Consequently, $I_\star^{\rm px}=0$ inside the planetary radius and $\mathcal{P}_{\rm s}^{\rm px}(x)$ cannot contribute to the photometric OSE in the planetary shadow. At any given observation time $t$, the sum

\begin{equation}\label{eq:F_OSE}
F_{\rm OSE}^1(t) = \left(\frac{R_{\rm s}}{R_\star}\right)^2 \ \sum_{\rm px} dx \times I_\star^{\rm px} \times P_{\rm s}^{\rm px}(x)
\end{equation}

\noindent
over all occulted pixels within the stellar radius ($R_\star$) gives us the relative loss in stellar brightness $B_{\rm OSE}^1(t)=1-F_{\rm OSE}^1(t)$ due to the photometric OSE of the first satellite. In the more general case of $n$ satellites,

\begin{equation}\label{eq:general_OSE}
F_{\rm OSE}^n(t) = \sum_{\rm s}^n {\Bigg [} \left(\frac{R_{\rm s}}{R_\star}\right)^2 \ \sum_{\rm px} dx \times I_\star^{\rm px} \times P_{\rm s}^{\rm px}(x) {\Bigg ]}
\end{equation}

\noindent
and $B_{\rm OSE}^n(t)=1-F_{\rm OSE}^n(t)$. We apply the nonlinear limb darkening law of \citet[][Equation~(7) therein]{2000A&A...363.1081C} and use stellar limb darkening coefficients (LDCs) listed in \citet{2011A&A...529A..75C}, which depend on stellar effective temperature ($T_{\star,{\rm eff}}$), metallicity ([Fe/H]), and surface gravity ($\log(g)$).

As an example, Figure~\ref{fig:limb_255.01} shows the simulated transit of KOI\,255.01 together with the OSE of a hypothetical Ganymede-sized moon at $a=15.47\,R_{\rm p}$, equivalent to Ganymede around Jupiter. The $2.51\,R_\oplus$ super-Earth ($R_\oplus$ being the Earth's radius) is an interesting object as it transits a $0.53\,M_\odot$-mass M dwarf with a radius of $0.51\pm0.06\,R_\odot$ every $27.52197994\pm3.295\times10^{-5}$\,d.\footnote{All values taken from \href{http://exoplanetarchive.ipac.caltech.edu}{http://exoplanetarchive.ipac.caltech.edu} as of 2014 May 30.} Hence, the photometric OSE of even a Ganymede-sized moon could be significant. In our simulations, the transit impact parameter is set to $\mathfrak{b}=0.0$ to ease comparison with the analytic model, although it is really given as $\mathfrak{b}=0.1244\,(+0.2132, -0.1243)$ in the Exoplanet Archive. The LDCs are $a_1=0.4354$, $a_2=0.2910$, and $a_3=0=a_4$. In both panels of Figure~\ref{fig:limb_255.01}, the solid line refers to our dynamical OSE model with limb darkening, the dashed line shows the dynamical model with the limb darkening option switched off, and the dotted line shows the analytical OSE model by \citet{2014ApJ...787...14H}, which also neglects limb darkening.

The left panel shows that the analytical solution is much less accurate than the dynamical OSE model inside the planetary transit trough because it neglects stellar limb darkening. In the wings of the transit curve, however, the analytical solution without stellar limb darkening and the dynamic OSE model with limb darkening differ by $<3\times10^{-6}$ compared to a maximum OSE depth of about $3\times10^{-5}$ just before the planetary ingress (right panel). The analytic model might thus offer sufficient accuracy for an initial OSE survey of a large data set. With such an approach, a first and preliminary search for OSE candidates within thousands of phase-folded light curves would be a matter of minutes. Even more encouraging, the dynamic OSE model with limb darkening almost resembles a straight line, at least in this configuration where the moon's semimajor axis is roughly as wide as the stellar diameter. A straight line fit would, of course, simplify an initial OSE search even further, as it would mostly depend on the moon's radius squared (in terms of maximum depth) and on the moon's semimajor axis (in terms of duration).

The right panel of Figure~\ref{fig:limb_255.01} also reveals that the dynamical OSE model causes a slightly smaller depth in the light curve for about two-thirds into the OSE ingress. This effect is caused by stellar limb darkening and the stellar brightness in the occulted regions being lower than the average brightness on the disk. In the final third of the OSE ingress, the dynamical OSE model then yields a deeper absorption because of the increasing stellar brightness towards the disk center. This division into two-thirds, in which the OSE model without stellar limb darkening is deeper than the one with limb darkening, and the one-third where things are reversed, is not a universal relation. Toward the stellar limb, e.g., the model without stellar limb darkening produces a deeper OSE signal during the entire transit.

\begin{figure*}[t]
   \centering
   \scalebox{0.46}{\includegraphics{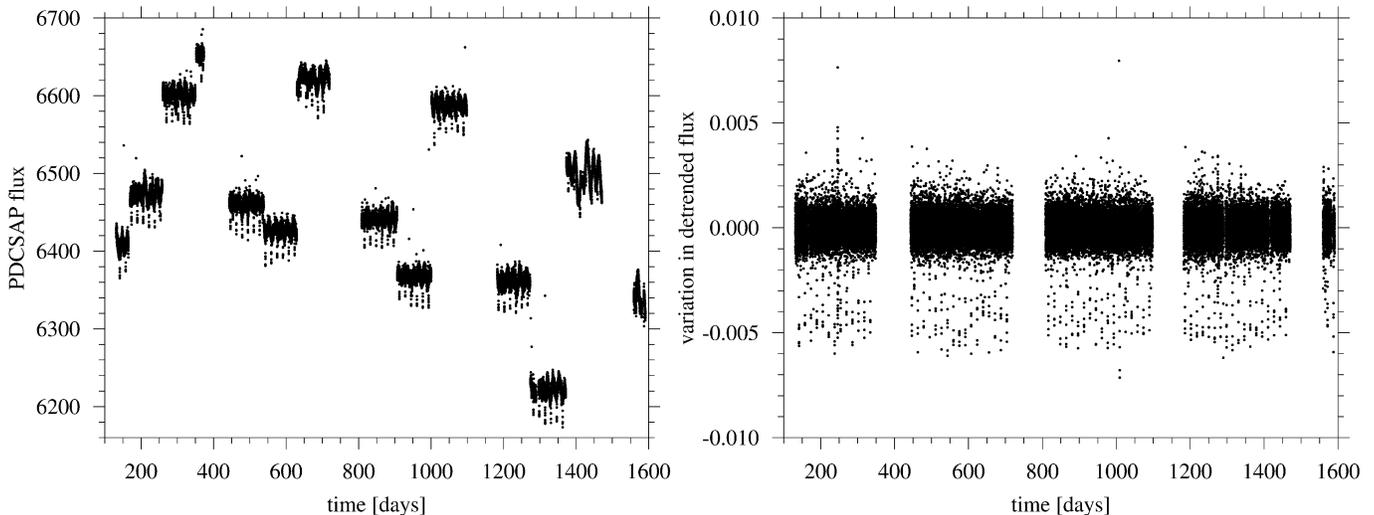}}
\caption{\textit{Left:} Raw PDCSAP\_FLUX for Kepler-229\,c -- individual quarters come offset from one another, and the approximately 90 transits are visible as large dips. \textit{Right:} Detrended data.}
\label{fig:condition_data_example}
\end{figure*}

\subsubsection{Comparison with \textit{Kepler} Data}
\label{subsub:comparison}

We now apply our model to observations. This investigation is not an in-depth search for moons around a test planet, but it shall serve as an illustration of the OSE by reference to actual observed data. Our survey for confirmed, super-Neptune-sized \textit{Kepler} planets with orbital periods $>10$\,d (to ensure Hill stability of an moons) around a $\gtrsim0.7\,R_\odot$ star (allowing detection of exomoons akin to the largest solar system moons) revealed Kepler-229\,c (KOI\,757.01) as a promising object. It is a $4.8\,R_\oplus$ planet transiting a $0.7\,R_\odot$ star every $16.96862$\,d with an impact parameter $\mathfrak{b}=0.25$ at a distance of about $0.117$\,AU \citep{2014ApJ...784...45R}. The stellar mass can then be estimated via Kepler's third law as $0.74\,M_\odot$, and the planetary Hill radius ($R_{\rm H}$) is about $78\,R_\oplus\approx16\,R_{\rm p}$, assuming that Kepler-229\,c's mass is similar to that of Neptune. Stellar LDCs are interpolated from \citet{2011A&A...529A..75C} tables using the stellar effective temperature and surface gravity of Kepler-229 \citep{2014ApJ...784...45R}, yielding $a_1=0.618423$, $a_2=-0.522542$, $a_3=1.24769$, and $a_4=-0.536854$.

We first retrieved as many quarters as were available among Q0-Q17 of the long-cadence\footnote{The OSE itself is an averaging effect, as it appears in the phase-folded light curve and only after several transits. Hence, for the OSE curve it does not make a difference if the data is taken in short cadence and then binned into 30\,minute intervals or if it is taken in 30\,minute intervals in the first place.} (30 minutes) publicly available \kepler~data for Kepler-229\,c.\footnote{\href{http://archive.stsci.edu/kepler/data\_search/search.php}{http://archive.stsci.edu/kepler/data\_search/search.php}} We analyzed the PDCSAP\_FLUX data, from which the \kepler\ mission has attempted to remove instrumental variability. Nevertheless, these data still exhibit significant variability unrelated to transits, as seen in Figure \ref{fig:condition_data_example} (left panel). The creation of the PDCSAP fluxes by the  \kepler\ mission involves the removal of common mode variability from the light curves attributable to instrumental effects, which can distort real astrophysical (such as stellar) variability but primarily at medium to long timescales. Since we consider relatively short-period planets, these possible distortions are unlikely to affect our analysis.

\begin{figure*}[t]
   \centering
   \includegraphics[angle=+90, width=\linewidth]{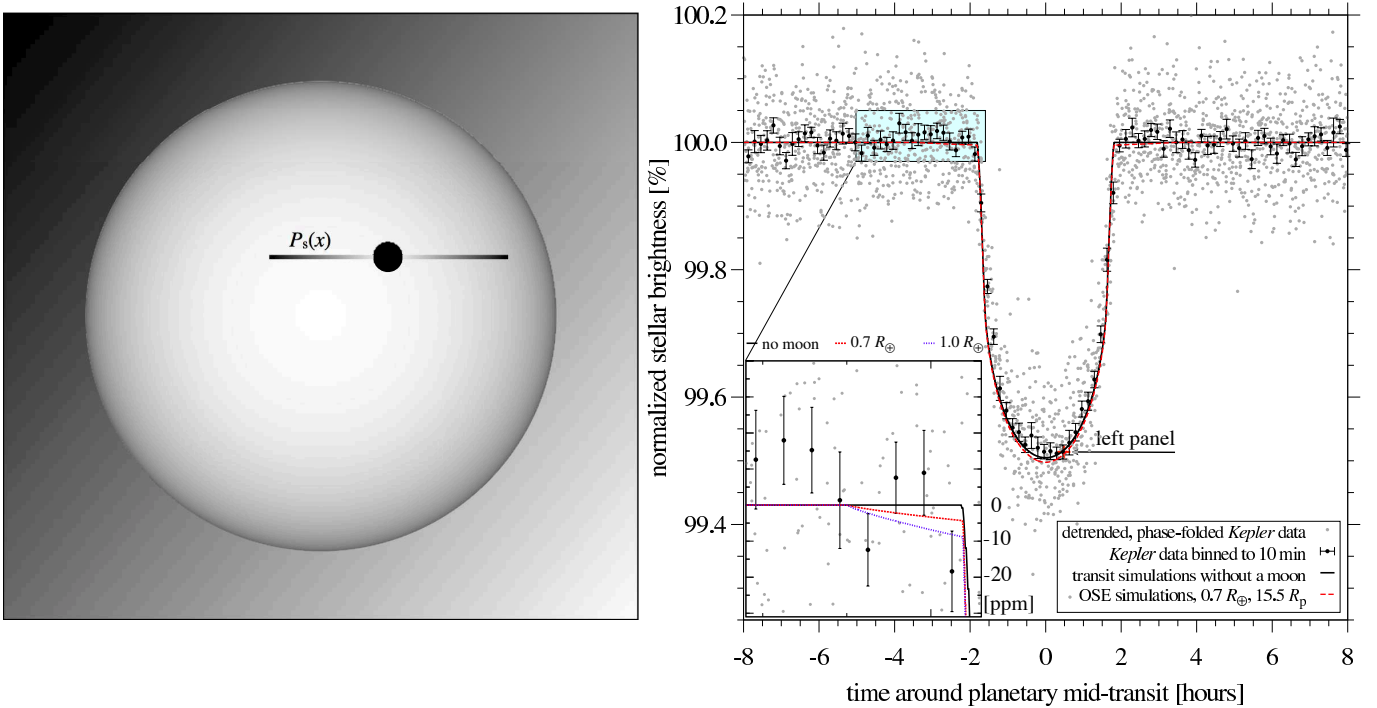}
   \caption{Visualization of our dynamical photometric OSE transit model. This example shows the transit of Kepler-229\,c, a $4.8\,R_\oplus$ planet orbiting a $0.7\,R_\odot$ star every $\approx17$\,d with a transit impact parameter $\mathfrak{b}=0.25$. \textit{Left:} The planet (black circle) and the probability function (shaded horizontal strip) of a hypothetical $0.7\,R_\oplus$ moon with an orbital semimajor axis of $8\,R_{\rm p}$ transit the limb-darkened star (large bright circle). \textit{Right:} The red cross on the OSE curve at 0.5\,hr refers to the moment shown in the left panel. The inset zooms into the wings of the transit ingress. Three models are shown: no moon (black solid), a $0.7\,R_\oplus$ moon (red dashed), and a $1\,R_\oplus$ moon (blue dotted), all moons with a semimajor axis of $8\,R_{\rm p}$ around the planet.}
   \label{fig:OSE}
\end{figure*}

To condition each quarter's observations, we subtracted the quarter's mean value from all data points and then normalized by that mean. To these mean-subtracted, mean-normalized data, we applied a mean boxcar filter with a width equal to twice the transit duration plus 10\,hr. This window is chosen to maximally remove non-transit variations while preserving the shape of the transit and OSE signals. The right panel of Figure \ref{fig:condition_data_example} shows the resulting detrended data. Finally, we stitched together all quarters and masked out 10$\sigma$ outliers.\footnote{We define $\sigma$ to be the standard deviation estimated as 1.4826 $\times$ the median absolute deviation \citep{2003drea.book.....B}.} Figure \ref{fig:OSE} (right panel) shows the final result for Kepler-229\,c, after we folded the detrended data on a period of \kerplertwotwoninecperiod. Gray dots present the detrended, phase-folded data, and black dots with error bars show the binned data. The solid black line shows the direct output of our dynamical OSE simulator but for a planet without a moon, and the dashed red line shows the transit including the photometric OSE of an injected moon. A red cross on that curve at about 0.5\,hr highlights the orbital configuration that is depicted in the left panel.

The left panel of Figure~\ref{fig:OSE} illustrates our dynamical OSE model at work for Kepler-229\,c. The planet along with the probability distribution of one hypothetical exomoon can be seen in transit. The injected moon has a radius of $0.7\,R_\oplus$, the vertical width of the moons $\mathcal{P}_{\rm s}(x)$ is to scale to both the planetary and the stellar radius. The moon orbit is set to $8\,R_{\rm p}$, which is $R_{\rm H}/2$ and therefore at the boundary of Hill stability for prograde moons \citep{2006MNRAS.373.1227D}.

The inset in the right panel zooms into the transit light curve just about an hour prior to planetary ingress. In addition to the black solid (no moon) and red dashed lines ($0.7\,R_\oplus$-sized moon), we also show a blue dotted line indicating the OSE of a hypothetical Earth-sized moon. Note that the width of individual error bars of the binned data is $\approx10^{-4}$, which corresponds to the depth of an Earth-sized moon's photometric OSE. With about 5 of such binned data points (or about 500 unbinned data points during the OSE ingress), a search for the exomoon-induced photometric OSE around this planet could yield statistical constraints on the presence of moons the size of Earth and smaller around this particular exoplanet. A more elaborate statistical analysis of exomoon effects in transit light curves is deferred to a future study (R. Heller et al. 2016, in preparation).

\section{Numerical Simulations of the Photometric OSE}
\label{sec:numerical}

To simulate the OSE in more complex configurations with inclined moon orbits, for which an analytical solution is not available, we wrote a numerical OSE simulator in {\tt python}, which we call ``{\tt PyOSE}''. The code and examples are publicly available\footnote{\href{https://github.com/hippke/pyose}{https://github.com/hippke/pyose}} under the MIT license.\footnote{\href{http://opensource.org/licenses/MIT}{http://opensource.org/licenses/MIT}} All of the following figures were generated with this code.

\subsection{Parameterization in {\tt PyOSE}}

\begin{figure}[t]
 \centering
  \includegraphics[angle=  0, width=0.97\linewidth]{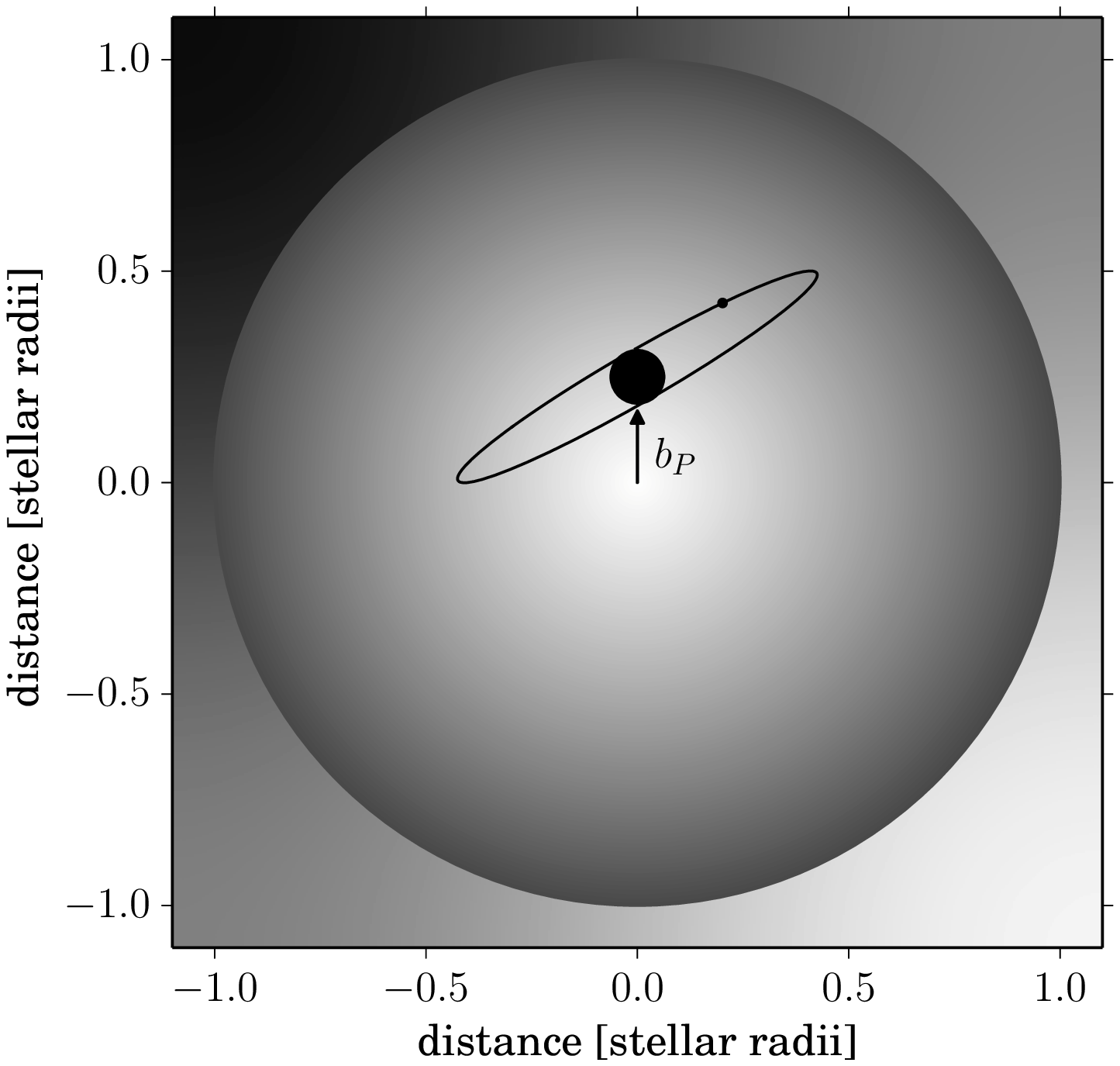}\\
  \includegraphics[angle=  0, width=0.96\linewidth]{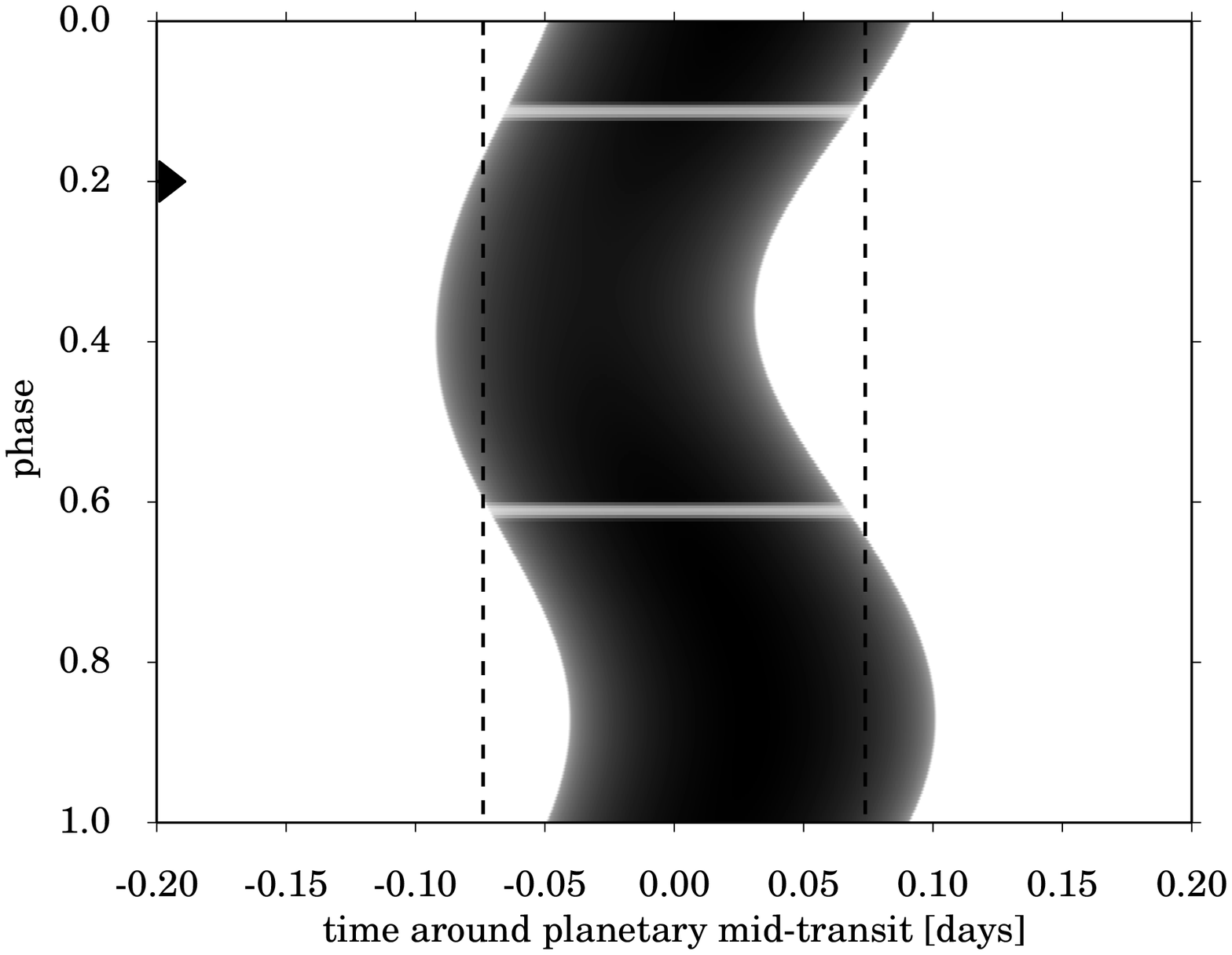}\\
  \includegraphics[angle=-90, width=0.96\linewidth]{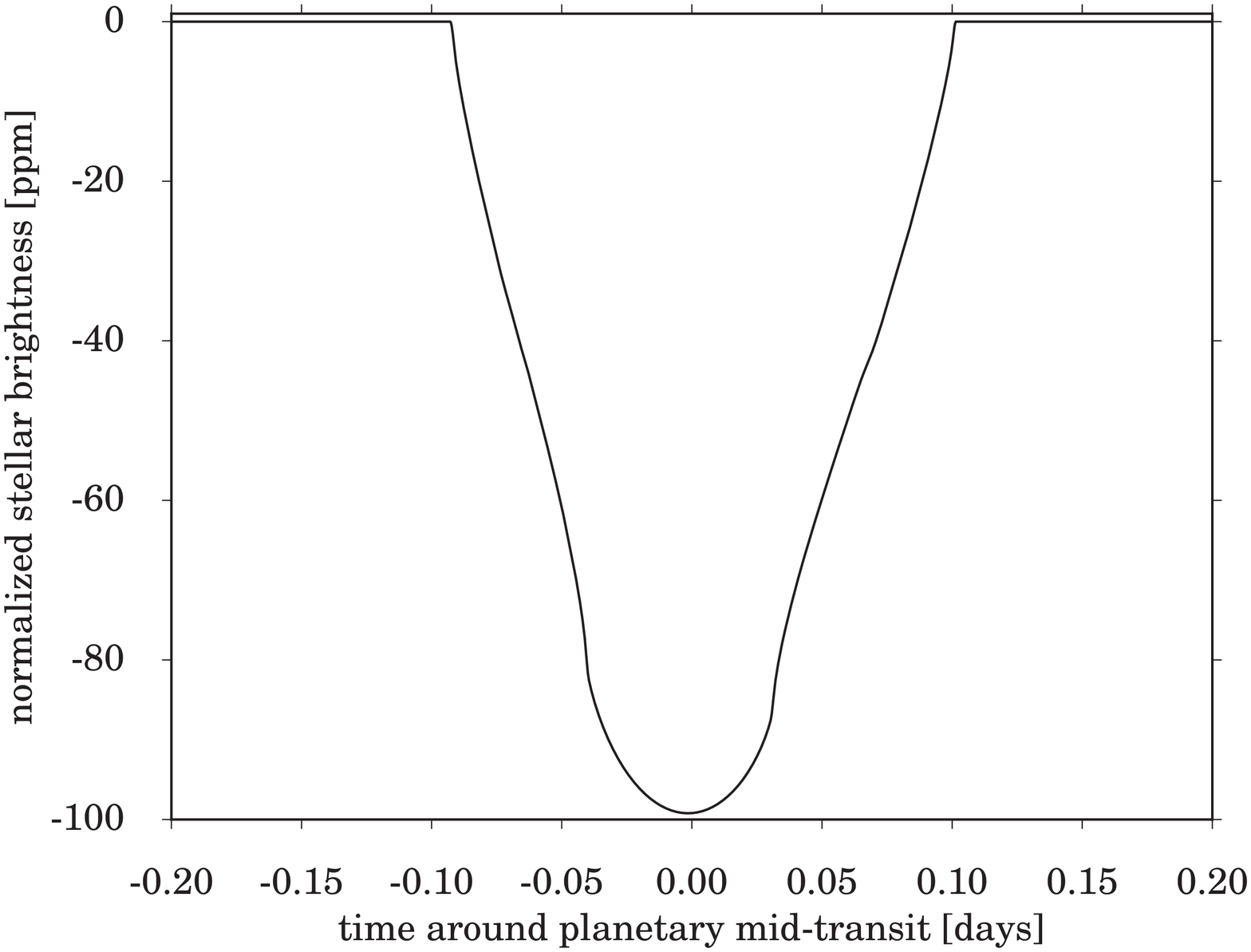}
  \caption{\textit{Top:} Sky-projected view generated with {\tt PyOSE} of both Kepler-229\,c and a hypothetical exomoon in transit. \textit{Center:} River plot of the moon transit only. The black arrow at phase 0.2 shows the moon's position chosen in the top panel. \textit{Bottom:} Average (phase-folded) transit light curve of the system after an arbitrarily large number of transits, where the moon orbit has been equally sampled in time.}
  \label{fig:code}
 \end{figure}

In {\tt PyOSE}, the moon's orbital ellipse is defined by its eccentricity ($e$), circumplanetary semimajor axis ($a$), its orbital inclination with respect to the circumstellar orbit ($i_{\rm s}$), the longitude of the ascending node ($\Omega$), and the argument of the periapsis ($\omega$). Figure~\ref{fig:code} (top) shows a hypothetical $0.7\,R_\oplus$ exomoon around Kepler-229\,c on a circular, inclined orbit ($i_{\rm s}=83^\circ$, $\Omega=30^\circ$, $a=8\,R_{\rm p}$) at the time of the planetary mid-transit. In our numerical implementation, the planet--moon ensemble transits the star from left to right, which is an arbitrary choice. The motion of the planet and the moon around their common barycenter during transit can be simulated with {\tt PyOSE}.

The center panel in Figure~\ref{fig:code} shows a river plot \citep{2012Sci...337..556C} representation of 250 of these transit light curves, with one shown in each row. Each of these light curves corresponds to a different fixed position of the moon during the stellar transit. The orbital phase of the moon (along the ordinate in Figure~\ref{fig:code}) corresponds to the mean anomaly. Two horizontal gray regions at phases $\approx0.1$ and $\approx0.6$ show partial planet--moon eclipses. The black arrow at phase $\approx0.2$ indicates the moon's position chosen in the top panel. For reference, the two vertical dashed lines illustrate the time of the planetary transit, which is omitted in our OSE simulations for clarity.

The bottom panel of Figure~\ref{fig:code} shows a sum of the individual moon transits from the river plot, normalized by the number of transits. This procedure corresponds to a phase-folding of a hypothetical observed OSE.

Although {\tt PyOSE} can simulate transits of planets with moons, we will focus on the moon's OSE signature in the following and study a range of moon transits for various orbital parameters. The transit model is the one presented by \citet{2002ApJ...580L.171M}. For each OSE curve, we chose to sample at least 100 moon transits equally spaced in time (not necessarily in space for $e\neq0$; see Figure~\ref{fig:sampling}) to achieve convergence \citep{2014ApJ...787...14H}. The limb-darkened stellar disk is represented by a numerical grid of $1\,000\times1\,000$ floating point values, and we calculate the total stellar brightness during transit for $\gtrsim 1\,000$ flux data points between first and last contact of the planetary silhouette with the stellar disk. With the moon having a radius of typically 1/100 the stellar radius in the shown simulations, or about 10\,px, its transiting silhouette is represented by about 100 black pixels. The error of this approximation compared to a genuine black circle is $<1$\,\%. {\tt PyOSE} allows for arbitrarily large pixel grids at the cost of CPU time, which is proportional to the total number of pixels in the grid or to the square of $R_{\rm s}$ (in units of pixels).

\begin{figure*}
\centering
\includegraphics[angle=-90,width=0.46\linewidth]{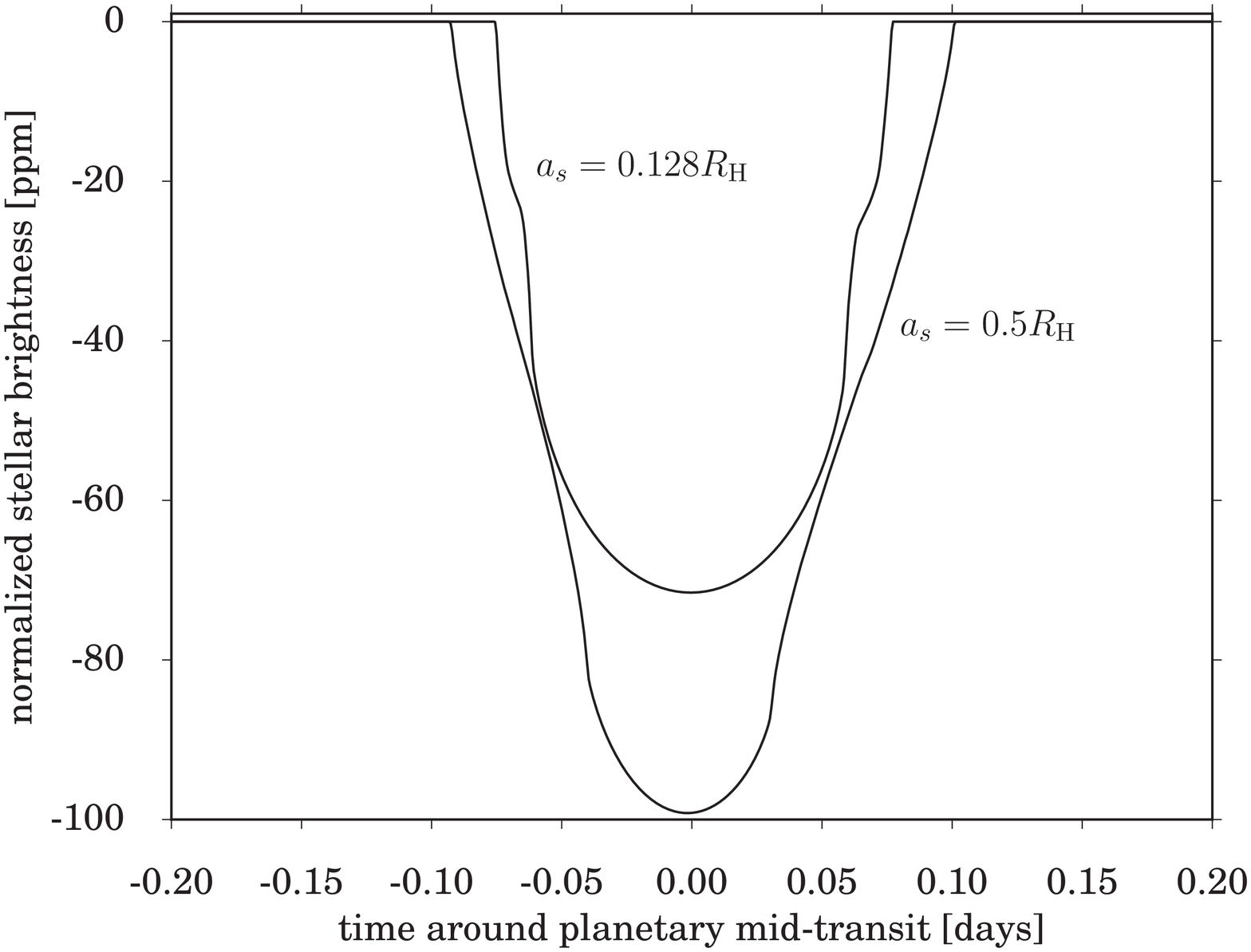}
\hspace{0.2cm}
\includegraphics[angle=-90,width=0.505\linewidth]{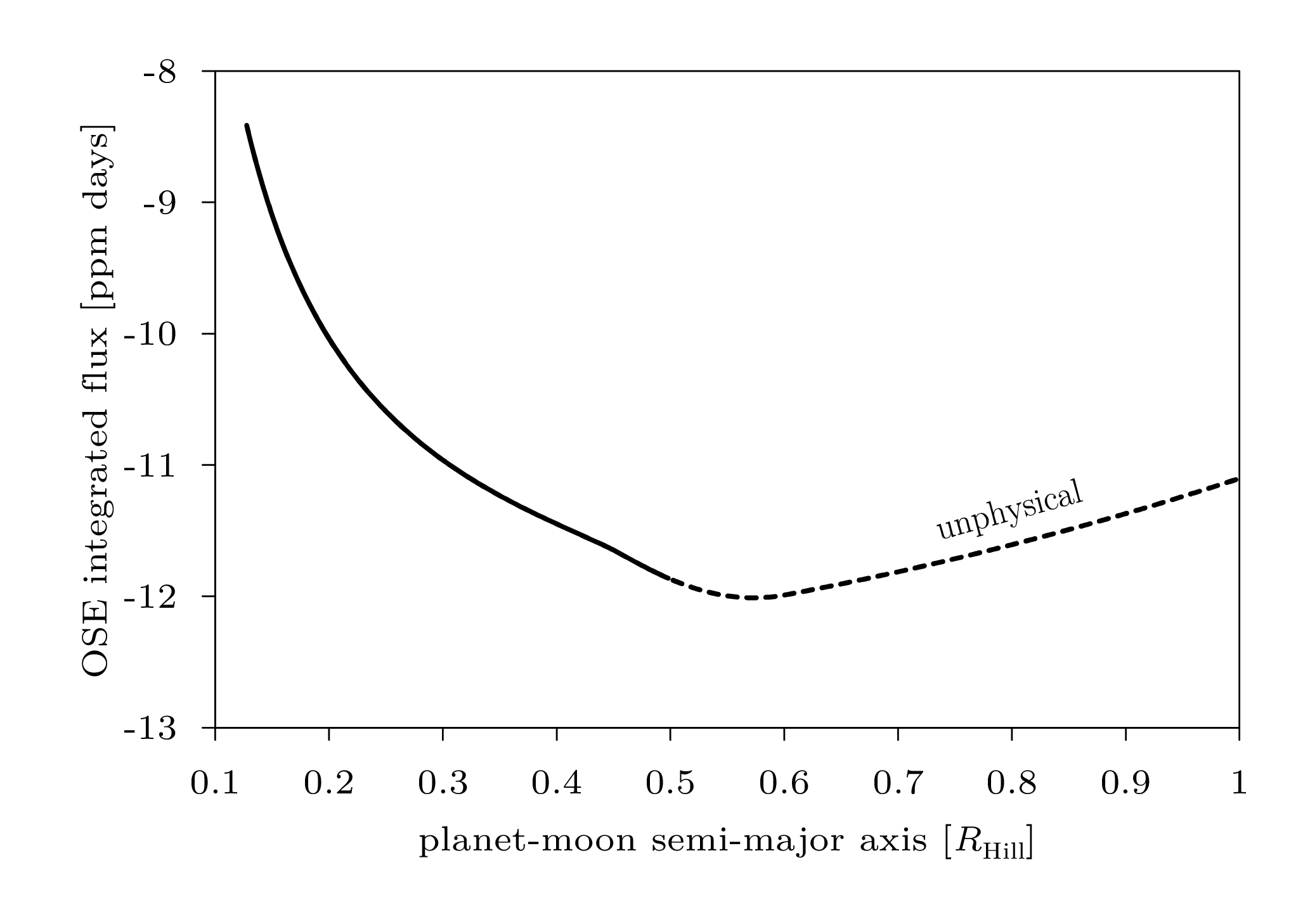}
\caption{Variation of the OSE for different semimajor axes of a hypothetical exomoon around Kepler-229c. \textit{Left:} Photometric OSE for two cases where the satellite is at $0.128$ and $0.5\,R_{\rm H}$ around the planet. \textit{Right:} Integral under the OSE curve as a function of the planet--moon orbital semimajor axis. The solid line shows the limiting cases at $0.128$ and $0.5\,R_{\rm H}$, corresponding to the two scenarios shown in the left panel. The dashed line represents moon orbits that are unphysically wide for prograde moons.}
\label{fig:axis}
\end{figure*}

\subsection{Mutual Planet--Moon Eclipses}

We treat both the planet and the moon as black disks. Thus, it is irrelevant whether the satellite eclipses behind or in front of the planet.\footnote{This fact makes the OSE insensitive to the satellite's sense of orbital motion around the planet \citep{2014ApJ...796L...1H,2014ApJ...791L..26L}.} For each of the simulated planet--moon transit configurations, {\tt PyOSE} compares the distance between the planet and moon disk centers ($d_{\rm ps}$) to the sum of their radii. If $d_{\rm ps}<R_{\rm p}+R_{\rm s}$, a mutual eclipse occurs and {\tt PyOSE} calculates the area ($A$) of the asymmetric lens defined by the intersection of the two circles as

\begin{align} \nonumber
A =& \ R_{\rm s}^2 \arccos\left(\frac{d_{\rm ps}^2 + R_{\rm s}^2 - R_{\rm p}^2}{2d_{\rm ps}R_{\rm s}}\right) \\ \nonumber
     &+ R_{\rm p}^2 \arccos\left(\frac{d_{\rm ps}^2 + R_{\rm p}^2 - R_{\rm s}^2}{2d_{\rm ps}R_{\rm p}}\right) \\ \nonumber
    &- {\Bigg (} \frac{1}{2} \sqrt{(-d_{\rm ps} + R_{\rm s} + R_{\rm p})(d_{\rm ps}+R_{\rm s}-R_{\rm p})} \\
    & \times \sqrt{(d-R_{\rm s}+R_{\rm p})(d_{\rm ps}+R_{\rm s}+R_{\rm p})} {\Bigg )}
\end{align}

\noindent
This area does not contribute to the stellar blocking by the moon, as it is covered by the silhouette of the transiting planet.

\subsection{Parameter Study}

In the following, we study variations of the OSE signal due to variations in the parameterization of the star--planet--moon orbital and physical configuration. We use Kepler-229\,c as a reference case and modify one parameter at a time, as specified below.

\subsubsection{The Moon's Semimajor Axis ($a$)}
\label{sub:axis}

For circular moon orbits, changes in $a$ modify the shape of the OSE, while the area under the integral remains unaffected (see Equation~\ref{eq:norm}). For $e\neq0$ or $i_{\rm s}\neq0$, however, both the shape and the integral will change. The left panel of Figure~\ref{fig:axis} visualizes this effect for two cases; one in which the moon is barely stable from a dynamical point of view ($a=0.5\,R_{\rm H}$), and one for a close-in moon near the Roche limit ($a\approx0.128\,R_{\rm H}\approx2\,R_{\rm p}$). The duration of the OSE signal is longer for moons with larger semimajor axes. Planet--moon eclipses are visualized by bumps in the OSE curve of the moon at $a=0.128\,R_{\rm H}$, near $\pm0.07$\,days. The moon in the wider orbit is not subject to planet--moon eclipses.

The right panel of Figure~\ref{fig:axis} shows the integral under the OSE curve as a function of $a$. This is an important quantity as it serves as a measure for the significance of the moon-induced OSE imprint on the light curve. The decline of the OSE integral up to $a\approx0.6\,R_{\rm H}$ is mostly due to stellar limb darkening: the larger $a$, the more flux will be blocked closer to the stellar center for this specific orbital configuration. Beyond $0.6\,R_{\rm H}$, moon transits will occasionally be missed during planetary transits and the OSE signal decreases. The dashed part of the curve, corresponding to moons beyond $0.5\,R_{\rm H}$, is physically implausible for prograde moons, and valid only for prograde moons \citep{2006MNRAS.373.1227D}.

\begin{figure}
\includegraphics[angle=-90,width=\linewidth]{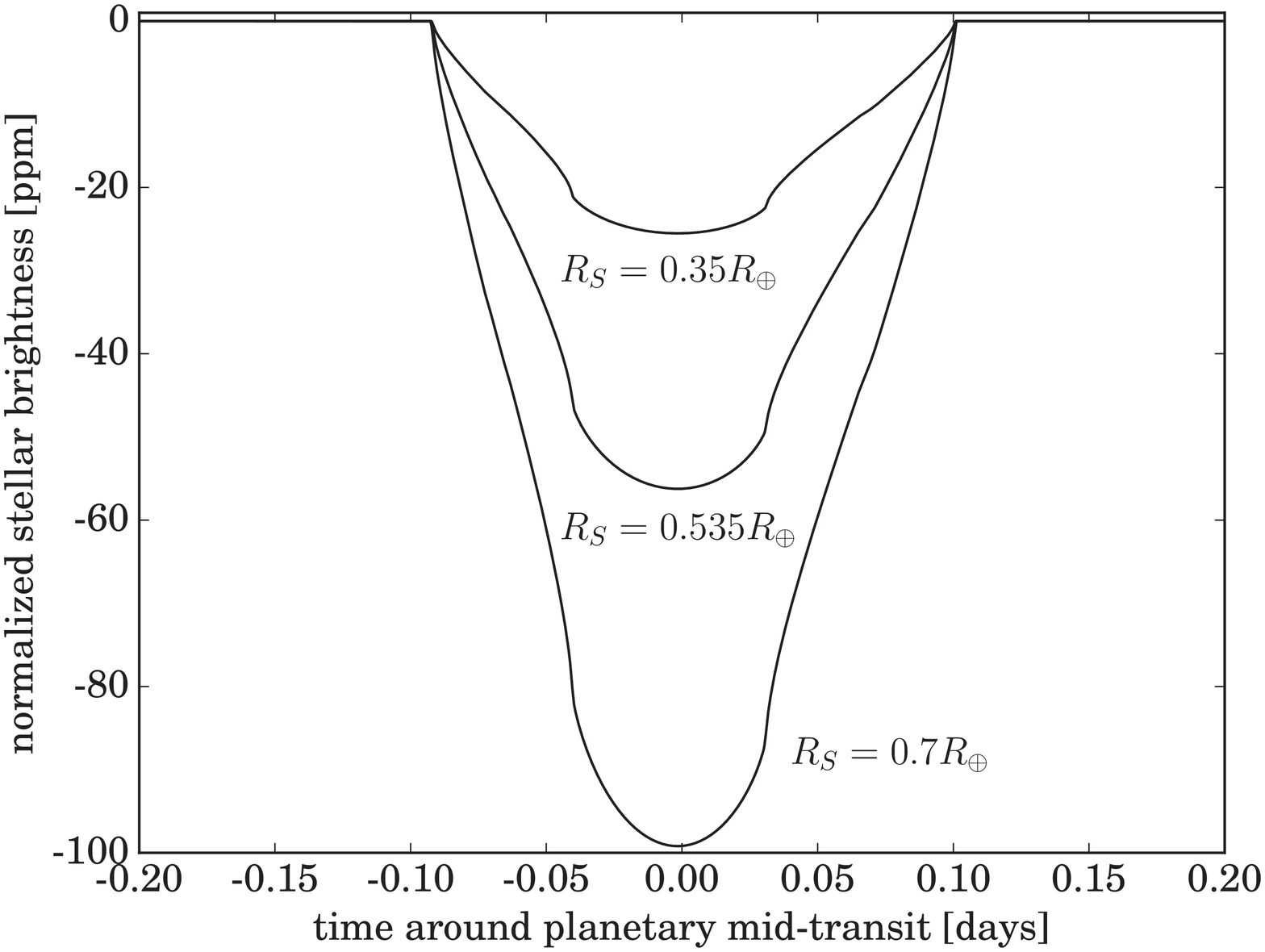}
\caption{Variation of the OSE for different satellite radii. The inset zooms into the ingress of the phase-folded satellite transit, showing that the time of first contact between the satellite's silhouette and the stellar disk depends on the satellite radius.}
\label{fig:radius}
\end{figure}

\subsubsection{The Moon's Radius ($R_{\rm s}$)}
\label{sub:radius}

Larger moons naturally cause deeper transits. Keeping everything else fixed, variations in $R_{\rm s}$ cause variations in the OSE amplitude roughly proportional to $R_{\rm s}^2$, as illustrated in Figure~\ref{fig:radius} (see also the term $(R_{\rm s}/R_\star)^2$ in Equation~\ref{eq:F_OSE}). Comparing the upper and lower curves, we see a signal increase by a factor of four (-25\,ppm vs. -100\,ppm) for a change in $R_{\rm s}$ by a factor of two (from 0.35 to $0.7\,R_\oplus$). Changes in transit duration occur due to the different timings of the first and last contact of the planetary silhouette with the stellar disk (see inset in Figure~\ref{fig:radius}).

\subsubsection{The Planetary Impact Parameter ($\mathfrak{b}$) and\\ the Inclination of the Moon's Orbit ($i_{\rm s}$)}
\label{sub:moon_incl}

\begin{figure}
\includegraphics[angle=-90, width=\linewidth]{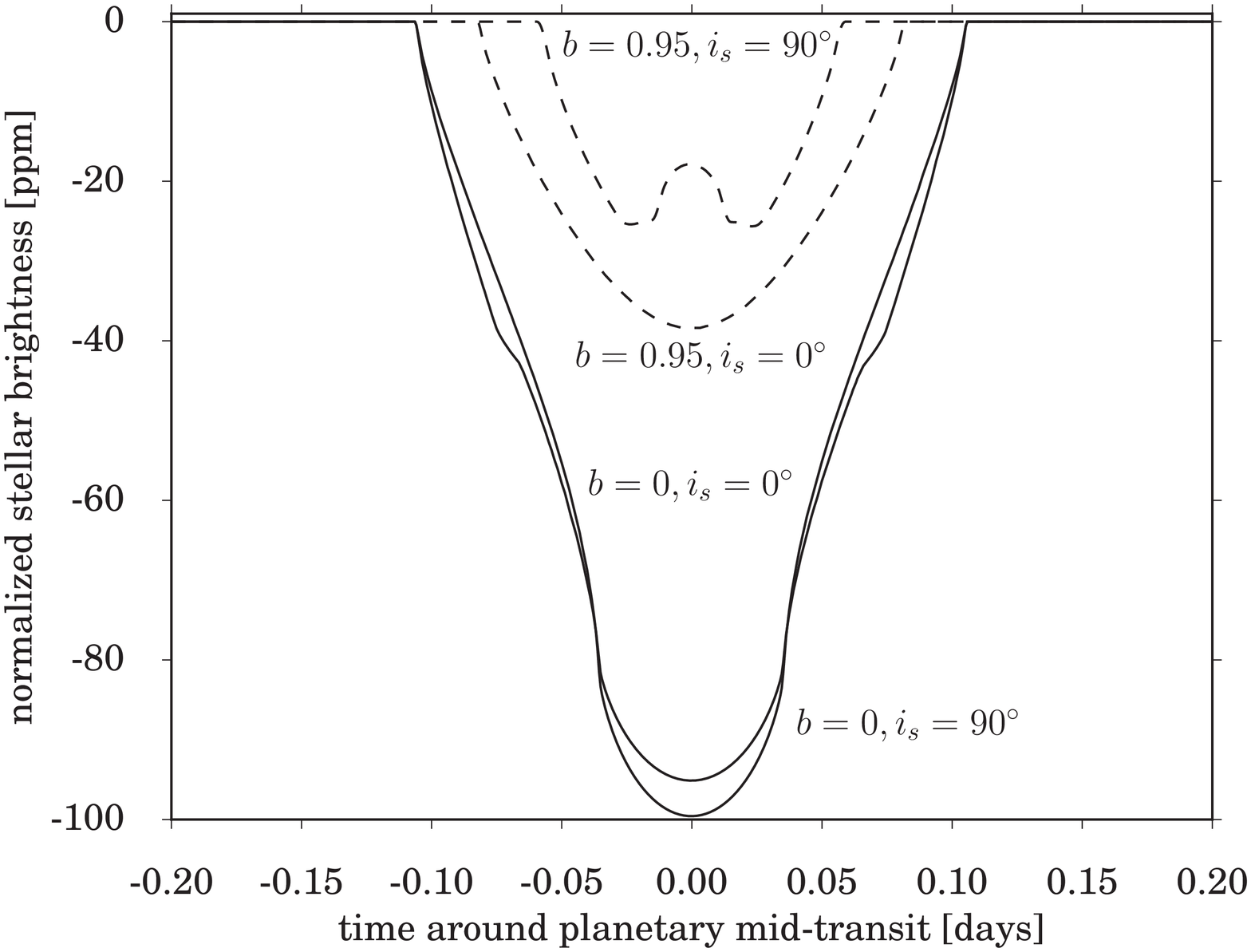}
\caption{Variation of the OSE for different planetary transit impact parameters and different inclinations of the satellite orbit. Solid lines refer to $\mathfrak{b}=0$ and dashed lines relate to $\mathfrak{b}=0.95$. For both cases, we show $i_{\rm s}=0^\circ$ and $90^\circ$.}
\label{fig:ib}
\end{figure}

The planetary impact parameter and the inclination of the satellite orbit determine the fraction of planetary transits without moon transits, that is, planetary transits with the moon passing beyond the stellar disk. In Figure~\ref{fig:ib}, we show four different scenarios of a hypothetical moon around Kepler-229\,c. Solid lines refer to $\mathfrak{b}=0$, dashed lines to $\mathfrak{b}=0.95$, and we examine inclinations $i_{\rm s}=0^\circ$ (face-on view) and $i_{\rm s}=90^\circ$ (edge-on view). In the $\mathfrak{b}=0.95$, $i_{\rm s}=90^\circ$ case (upper dashed line), a bump around planetary mid-transit gives evidence of planet--moon eclipses. In the $\mathfrak{b}=0.95$, $i_{\rm s}=0^\circ$ and $\mathfrak{b}=0$, $i_{\rm s}=0^\circ$ cases, planet--moon eclipses do not occur. In the $\mathfrak{b}=0$, $i_{\rm s}=90^\circ$ case (lower solid line), the bump from planet--moon eclipses is very broad and deformed into two minor bumps at $\pm0.07$\,d in the moon's OSE.

\begin{figure}
\includegraphics[angle=-90, width=\linewidth]{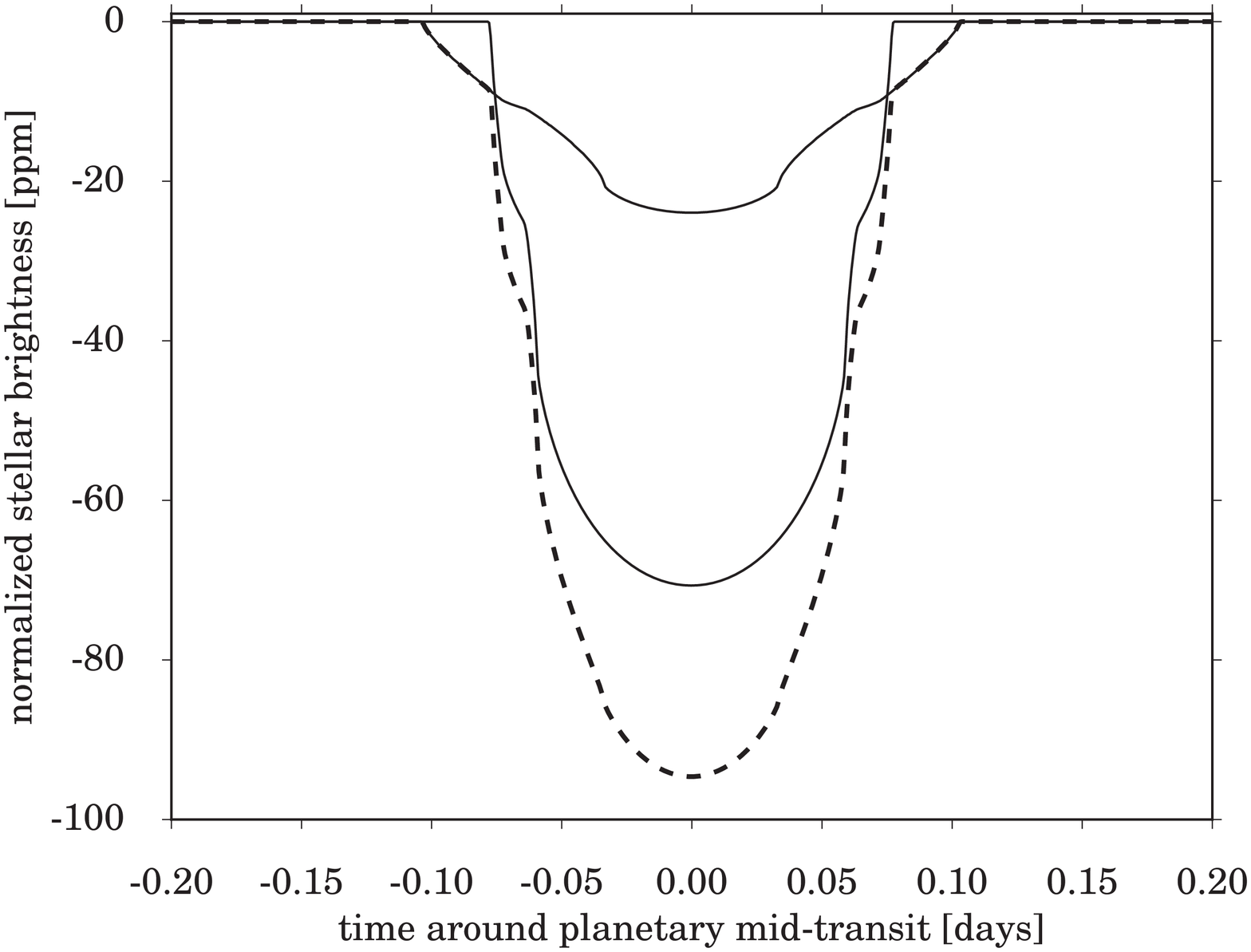}
\caption{OSE of a multi-moon system. The two solid lines show the OSE of one $0.35\,R_\oplus$ moon at $0.5\,R_{\rm H}$ and a second $0.7\,R_\oplus$ moon at $0.128\,R_{\rm H}$. The dashed line shows the combined OSE.}
\label{fig:multi-moon}
\end{figure}

\subsection{Multiple Exomoons}

Multiple moons are common in our solar system. Although photodynamical modeling \citep{2011MNRAS.416..689K} can tackle multi-satellite systems in principle \citep{2015ApJ...813...14K}, this is extremely time-consuming and a practicable approach still needs to be demonstrated. Following \citet[][Sect. 2.3 therein]{2014ApJ...784...28K}, the HEK team indeed restricts their exomoon search in the \textit{Kepler} data to single-moon systems. Our Equation~(\ref{eq:general_OSE}) is an analytical description of the additive OSE in multi-moon systems and describes how the dynamical OSE model handles multi-satellite systems. Our purely numerical simulator {\tt PyOSE} can handle multi-satellite systems as well.

That being said, with real observational data, distinguishing between an OSE signal of a single large moon and a signal from several smaller moons may prove difficult, if not impossible. This is demonstrated in Figure~\ref{fig:multi-moon}, where we show the individual OSE signals from two moons (solid lines) and their combined OSE signature (dashed line). This dashed curve demonstrates that the photometric OSE is additive as long as moon--moon occultations can be neglected. \footnote{{\tt PyOSE} currently neglects moon--moon occultations. They would only occur in $<1\,\%$ of the transits, depending on the exact geometry of the circumstellar and circumplanetary orbits.} The dashed curve is also what would be observed in reality, noise effects aside. With noise taking into account, there would be substantial degeneracies among the model parameters and the number of moons.

\section{Discussion}
\label{sec:discussion}

In addition to the orbital and physical parameters treated above, there are several minor effects on the OSE, the most important being stellar limb darkening. We ran a series of simulations using different LDCs for FGK main-sequence stars. For a typical uncertainty in stellar temperature $\approx$100K, the effect on the OSE signal is of the order of 1\,ppm, that is, negligible for the vast majority of observable cases. Nevertheless, if LDCs cannot be constrained otherwise, there could be degenerate solutions between an exomoon-induced OSE and a sole planetary transit with a different LDC parameterization. This is, however, only an issue during planetary transit. Alternatively, one can search for OSE-like flux decreases before and after the planetary transit \citep{2015ApJ...806...51H}, which cannot possibly be made up for with different LDCs.

Some aspects of our OSE model might hardly be accessible with near-future technology. For moon orbits that are co-planar with the circumstellar orbit ($i_{\rm s}=90^\circ$), moderate moon eccentricities will not cause an OSE signal much different from a circular moon orbit. Nevertheless, there are configurations in which these parameters make all the difference in determining the presence of an exomoon (see Fig.~\ref{fig:ib}). Moreover, the discovery of the formerly unpredicted hot Jupiter population, the unsuspected dominance of super-Earth-sized exoplanets in orbits as short as that of Mercury, and the puzzling abundance of close-in planets with highly misaligned orbits suggests that the solar system does not present a reliable reference for extrasolar planetary systems. Hence, any model used to explore yet undiscovered exomoons will need to be able to search a large parameter space beyond the margins suggested by the solar system.

Both numerical simulations and observations will always be undersampled and only converge to the analytical solution. In terms of observations, this is because of the limited number of observed transits, usually $<100$ for an observational campaign over a few years, and because of telescope downtime and observational windows. Beyond that, the orbital periods of both the planet--moon barycenter (around the star) and the moon(s) (around the planet--moon barycenter) \textit{determine} a non-randomized sampling of the moon orbit in successive transits. If an exoplanet's orbital period were -- by whatsoever reason -- an integer multiple of its moon's period, then the moon were to appear at the same position relative to the planet in successive transits. The resulting phase-folded light curve would not display the OSE and therefore not converge to our solutions, but it would show two transits: one caused by the planet and one caused by the moon.

{\tt PyOSE} can simulate large numbers of OSE curves for arbitrary star--planet--multi-moon configurations to test real observations. Beyond the functionalities demonstrated in this paper, {\tt PyOSE} can add noise (parameterized or injected real noise).

Both our dynamical OSE model (Sect.~\ref{subsub:dynamical}) and our numerical OSE simulator ({\tt PyOSE}, Sect.~\ref{sec:numerical}) are computationally inexpensive and easy to implement in computer code, which is crucial for the independent verification or rejection of possible exomoon signals. More advanced methods may suffer from a large parameter space to be explored, resulting in a huge number of simulations \citep[$10^{11}$;][]{2013ApJ...777..134K}. These could be difficult to verify. For an independent verification of an exomoon search, either the original code needs to be released for review, or an independent implementation is required. To exclude processing, runtime, and hardware errors (bit error rates are typically $10^{-14}$), a search based on TBs of data ultimately needs to be repeated on different hardware. The average computational burden for this is large, with an average of roughly 33,000 CPU hours required per candidate using photodynamical modeling. The monetary equivalent, e.g. using Amazon's EC2 on-demand facilities\footnote{\href{https://aws.amazon.com/ec2}{https://aws.amazon.com/ec2}}, is about \$50,000 U.S. dollars (2015 December prices) for a single candidate check.

The dynamical OSE simulation in Figure~\ref{fig:limb_255.01} contains 48 data points and was computed within 10 to 14\,s on two modern computers.\footnote{Computations took 14\,s on a MacBook Pro Retina 8-core, 2.8\,GHz Intel Core i7 processor, 16\,GB of total memory, 1600\,MHz DDR3 and 10\,s on a desktop computer with an 8-core, 3.6\,GHz Intel Core i7-3820 CPU, 32\,GB of total memory.} Hence, this setup can generate a grid of about $10^4$ such OSE light curves per day. Our OSE model involves 11 independent parameters: $M_\star$, $R_\star$, the planet's orbital period around the star ($P_{\rm {\star}p}$), $R_{\rm p}$, $\mathfrak{b}$, $a_1$, $a_2$, $a_3$, $a_4$, $R_{\rm s}$, and $a$. For a well parameterized star--planet system, $M_\star$, $R_\star$, $P_{\rm {\star}p}$, and $\mathfrak{b}$ can be observed and fit without considerations of any potential moon, assuming that moon-induced variations of the transit impact parameter \citep{2009MNRAS.396.1797K} are negligible. If one were to carry out a search for the photometric OSE in the \textit{Kepler} data, a limb darkening law with two LDCs should still do a good job in a first broad survey, leaving us with $R_{\rm p}$, $R_{\rm s}$, and $a$ plus the two LDCs to be fit per phase-folded light curve. Testing 10 values per parameter would then imply a grid of $10^5$ OSE light curves per planet or planet candidate. With $10^4$ LCs simulated per day, one \textit{Kepler} planet or planet candidate could be checked for an OSE signature within a week, given a standard desktop computer. If dedicated high-speed computational resources could be used, all \textit{Kepler} planets and candidates (about 4000 as of today) could be checked for a photometric OSE within maybe a month. A simple straight line fit of the OSE, as suggested above, would dramatically reduce this time frame to much less than one day. A detailed statistical analysis, e.g. within a Bayesian framework \citep{2012ApJ...750..115K} and using an injection-retrieval method \citep{2015ApJ...806...51H}, could then be used to infer the significance of the best-fit model for each object.

\section{Conclusion}
\label{sec:conclusion}

We present a new formula to describe the orbital sampling frequency of a moon on an eccentric orbit around a planet, that is, the probability of a moon residing at a specific sky-projected distance from the planet (Equation~\ref{eq:Ps_x_final}). This formula assumes co-planar circumstellar and circumplanetary orbits. We implemented it in a dynamical OSE simulator with stellar limb darkening that can be applied to arbitrary transit impact parameters. In contrast to a previously derived framework, in which stellar limb darkening was neglected and $\mathfrak{b}$ was required to be zero \citep{2014ApJ...787...14H}, our new dynamical OSE simulator can now be applied to observations (Figure~\ref{fig:OSE}).

Using an independent numerical OSE simulator dubbed {\tt PyOSE}, we examined the moon's part of the OSE parameter space, spanned by its orbital semimajor axis, its physical radius, the inclination of its orbit with respect to the line of sight, the orbit's longitude of the ascending node, and the transit impact parameter of the planet (and therefore of the moon).

The OSE might give evidence of a multi-moon configuration, but the precise characterization of multi-satellite system will be extremely challenging using OSE only. Therefore, the OSE method could be used for preliminary analyses of a large number of systems, while more costly methods \citep{2012ApJ...750..115K,2014ApJ...796L...1H,2015ApJ...812....5A} could be used to focus on the most promising subset of targets. Beyond that, the OSE method can generally be used as an independent means to verify an exomoon claim via planetary TTV and TDV.\\

\acknowledgments
We thank the referee for diligent reports. Ren{\'e} Heller has been supported by the Origins Institute at McMaster University, by the Canadian Astrobiology Program (a Collaborative Research and Training Experience Program funded by the Natural Sciences and Engineering Research Council of Canada), by the Institute for Astrophysics G\"ottingen, and by a Fellowship of the German Academic Exchange Service (DAAD). This work made use of NASA's ADS Bibliographic Services. Computations were performed with {\tt ipython 0.13} on {\tt python 2.7.2} \citep{PER-GRA:2007}, and figures were prepared with {\tt gnuplot 4.6} (\url{www.gnuplot.info}).

\bibliography{ms}
\bibliographystyle{apj2}

\end{document}